# Three-dimensional Hydrodynamic Bondi-Hoyle Accretion. IV.

## Specific Heat Ratio 4/3

**M. Ruffert**[*]

Max-Planck-Institut für Astrophysik, Postfach 1523, D - 85740 Garching, Germany

March 6, 1995

**Abstract.** We investigate the hydrodynamics of three-dimensional classical Bondi-Hoyle accretion. A totally absorbing sphere of different sizes (1, 0.1 and 0.02 accretion radii) exerts gravity on and moves at different Mach numbers (0.6, 1.4, 3.0 and 10) relative to a homogeneous and slightly perturbed medium, which is taken to be an ideal gas ($\gamma = 4/3$). We examine the influence of Mach number of the flow and size of the accretor upon the physical behaviour of the flow and the accretion rates.

The hydrodynamics is modeled by the "Piecewise Parabolic Method" (PPM). The resolution in the vicinity of the accretor is increased by multiply nesting several $32^3$-zone grids around the sphere, each finer grid being a factor of two smaller in zone dimension than the next coarser grid. This allows us to include a coarse model for the surface of the accretor (vacuum sphere) on the finest grid while at the same time evolving the gas on the coarser grids.

For small Mach numbers (0.6 and 1.4) the flow patterns tend towards a steady state, while in the case of supersonic flow (Mach 3 and 10) and small enough accretors, (radius of 0.1 and 0.02 accretion radii) an unstable Mach cone develops, destroying axisymmetry. Our 3D models do not show the highly dynamic flip-flop flow so prominent in 2D calculations performed by other authors. In the gamma=4/3 models, the shock front remains closer to the accretor and the mass accretion rates are higher than in the gamma=5/3 models, whereas the rms of the specific angular momentum accreted does not change.

## 1. Introduction

This paper extends our investigation (started in Ruffert, 1994a, Ruffert & Arnett, 1994 and Ruffert 1994b; henceforth referred to as papers I, II and III, respectively) of the classic Bondi-Hoyle-Lyttleton accretion model. In this scenario a totally absorbing sphere exerts gravity on and moves with velocity $v_\infty$ relative to a surrounding homogeneous medium of density $\rho_\infty$ and sound speed $c_\infty$. One would like to find the accretion rates of various quantities (mass, angular momentum, etc.) as well as the properties of the flow (e.g. distribution of matter and velocity, stability, etc.). In this fourth instalment we investigate what influence the adiabatic index $\gamma$ has on these properties and quantities, by repeating the simulations done in the previous three papers except for choosing $\gamma = 4/3$. For the Bondi-Hoyle accretion model, the value $\gamma = 5/3$, that was used throughout the previous papers, is a degenerate case: the distance to the sonic point, is given by

$$d_s = \frac{5-3\gamma}{4} R_B \quad, \tag{1}$$

with the Bondi radius (Bondi, 1952) given by

$$R_B = \frac{GM}{c_\infty^2} \quad. \tag{2}$$

The accretion radius (Hoyle & Lyttleton, 1939, 1940a, 1940b, 1940c; Bondi & Hoyle, 1944) is

$$R_A = \frac{2GM}{v_\infty^2} \quad, \tag{3}$$

in which $M$ is the mass of the accretor, $G$ the gravitational constant, $c_\infty$ the sound speed of the medium at infinity and $v_\infty$ the bulk flow of the medium at infinity. For a value of $\gamma$ of 5/3 the sonic distance is zero: $d_s(\gamma = 5/3) = 0$. So in all previous models (of papers II and III) the finite size of the accretor is always larger than the sonic distance. Whereas some previous work (Hunt, 1979; Petrich et al, 1989; the latter investigates relativistic accretion) has shown this difference to be important, other workers (e.g. Matsuda et al, 1991 and 1992) find fairly similar flow structures when varying $\gamma$. With our increased resolution we would like to investigate similarities and differences of the flow.

[*] e-mail: `mor@mpa-garching.mpg.de`

heat ratio $\gamma = 4/3$ is very similar to the flow of a gas with $\gamma = 5/3$. Except for one model (BS) the differences are quantitative in nature, not qualitative.

The aim of this paper is to more clearly delineate these differences, to compare the accretion rates of several quantities when flow parameters are changed and to obtain a more uniform picture of the accretion rates. In section 2 we give only a short summary of the numerical procedure used, since it has been described in the previous papers I, II and III. Section 3 presents the results, which we analyze and interpret in Sec. 4. Section 5 summarizes the implications of this work.

## 2. Numerical Procedure and Initial Conditions

Since the numerical procedures and initial conditions are mostly identical to what has already been described and used in papers I, II and III we will refrain from repeating every detail. Instead we only give a brief summary.

### 2.1. Numerical Procedure

A gravitating, totally absorbing "sphere" moves through an initially slightly perturbed (3%) homogeneous medium. Its velocity $v_\infty$ is varied in different models: we do simulations with Mach numbers $\mathcal{M}_\infty$ of 0.6, 1.4, 3.0 and 10. Matter flows in +x-direction. Our units are (1) the sound speed $c_\infty$ as velocity unit; (2) the accretion radius (Eq. 3) as unit of length, and (3) $\rho_\infty$ as density unit. Thus the unit of time is $R_A/c_\infty$.

The distribution of matter on multiply nested (e.g. Berger & Oliger, 1984) equidistant Cartesian grids with zone size $\delta$ is evolved using the "Piecewise Parabolic Method" (PPM) of Colella & Woodward (1984). The equation of state is that of a perfect gas with a specific heat ratio of $\gamma = 4/3$. The model of the maximally accreting, vacuum sphere in a softened gravitational potential is summarized in paper III.

The calculations are performed on a Cray-YMP 4/64 and a Cray EL. They need about 12 MWords of main memory and take approximately 40 YMP-CPU-hours per simulated time unit (for the $\delta = 1/64$ models and Mach 10; the $\delta = 1/256$ models take four times as long, etc.; $\delta$ is the size of a zone on the finest grid, see Table 1).

### 2.2. Models

The combination of parameters that we varied, together with some results, are summarised in Table 1. The first letter in the model designation indicates the Mach number: A, B, C and D stand for Mach 0.6, 1.4, 3.0 and 10, respectively. The second letter specifies the size of the accretor: L (large), M (medium) and S (small) stand for accretor radii of 1, 0.1, and 0.02 $R_A$, respectively.

Note that the L models might be physically somewhat unrealistic since they imply "giant" vacuum holes with an only weak gravitational field. However, one cannot rule out the case dius, e.g. in the accretion during nova outbursts (K. Schenker, 1995, private communication; Shara et al, 1986). Additionally, the models are very useful to obtain a general overview of the influence of the accretor size, which in turn allows a better judgement of the numerical quality of the simulations.

The size of the largest grid $L$ is adapted to the size of the accretor, larger accretors needing larger grids because the evolution time scales are longer and thus the gravitative influence reaches further out. The grids are nested to such a depth $g$ that the radius of the accretor $R$ spans several zones on the finest grid and the softening parameter $\epsilon$ is then chosen to be a few zones less than the number of zones that the accretor spans.

As far as computer resources permitted, we aimed at evolving the models for at least as long as it takes the flow to move from the boundary to the position of the accretor which is at the center (crossing time scale). This time is given by $L/2\mathcal{M}_\infty$ and ranges from about 1 to about 100 time units. The actual time $t_f$ that the model is run can be found in Table 1.

## 3. Description of Results

We will not only describe the results obtained in the new simulations, but also compare these results with those obtained in papers II and III for models with $\gamma = 5/3$.

### 3.1. Subsonic accretors

In models AS, AM and AL the relative velocity of the accretor to the surrounding medium is subsonic (Mach 0.6), and so no accretion shock cones are expected or seen in the simulations: Fig. 1 shows snapshots of the density distribution together with the flow pattern for all three models at the end of the simulation. The temporal evolution of the accretion rates (also displayed in Fig. 1) shows that a steady state develops after initial transients have died down. These transients result because the density distribution is initially randomly perturbed.

These plots can directly be compared to Figs. 1 and 2 of paper III. The overall flow pattern of corresponding models with the same Mach number is very similar, so also the pertaining comments of paper III apply here: radial inflow and spherical contours at distances shorter than 0.4 $R_A$, elliptical contours at larger distances. In the direct vicinity of the accretor (at distances of about 0.1 $R_A$ or less) the flow is practically spherically symmetric (model AS), while at larger distances (0.5 $R_A$ or more) the flow is is only mildly deflected by the weak gravitational field. A local density maximum is built up just downstream of the accretor. Although the absolute value of the mass accretion rate is slightly different in corresponding models (we will discuss this point in Sec. 4), the evolution in time is again very similar: the mass accretion rate tends toward a constant value while the angular momentum components converge toward zero after initial transients die down.

The values of some of the steady state accretion rates have been collected in Table 1. They will be compared to the values of models with $\gamma = 5/3$ in the sections that follow further below.

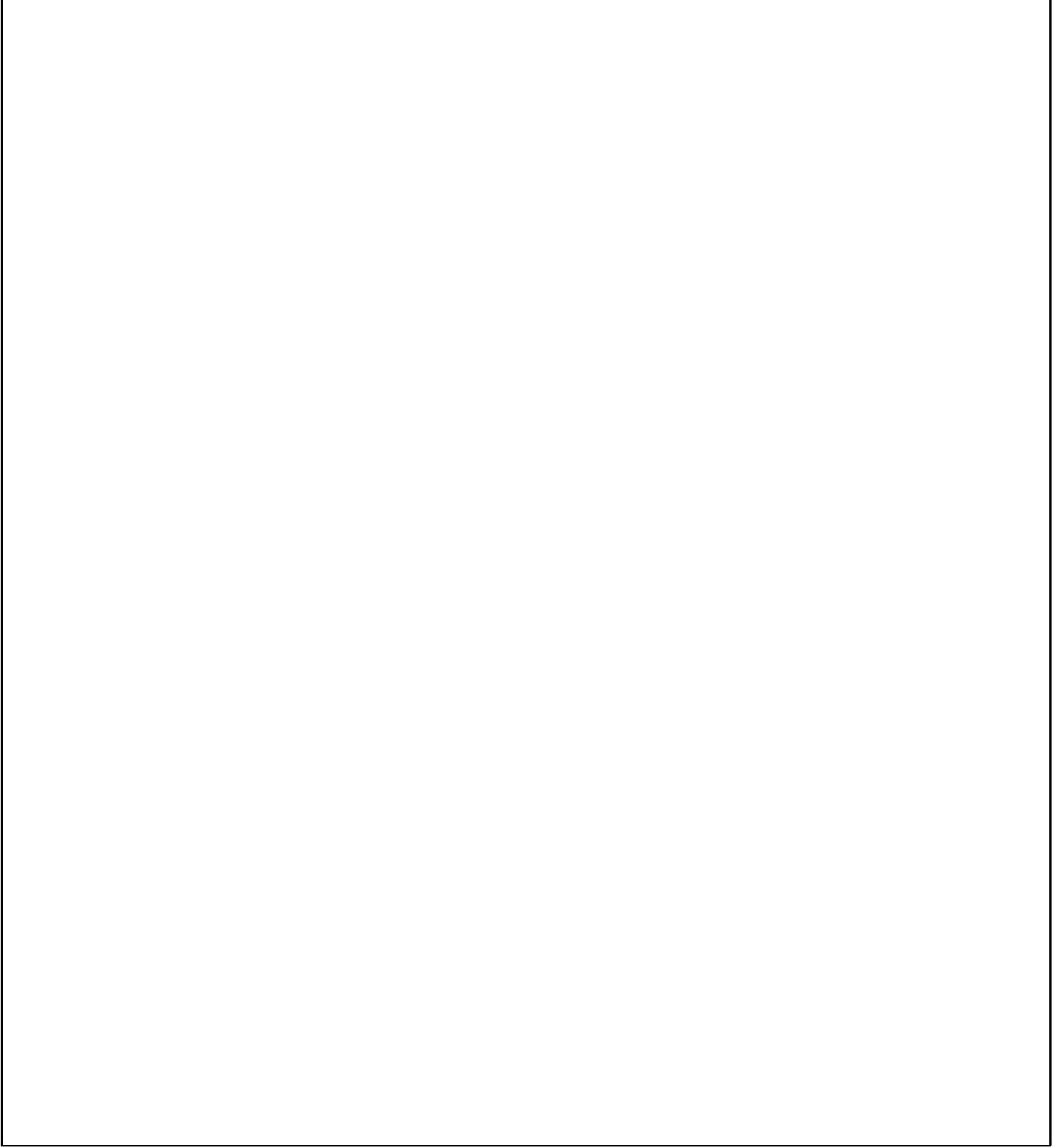

**Fig. 1.** Panels a, c and e are contour plots showing snapshots of the density together with the flow pattern in a plane containing the center of the accretor for all subsonic models AL (panel a), AM (panel c), and AS (panel e). The contour lines are spaced logarithmically in intervals of 0.1 dex for model AS and 0.05 dex for models AM and AL. The bold contour levels correspond to $\log \rho = 0.01$ for model AL and $\log \rho = 1$ for model AS, while the contours of model AM range from $\log \rho = 0.1$ to $0.45$ in increments of $0.05$. The time of the snapshot together with the velocity scale is given in the legend at the upper right hand corner in all three panels. The accretion rates of several quantities are plotted as a function of time for all subsonic models AL, AM and AS in panels b, d and f. The straight horizontal lines show the analytical mass accretion rates: dotted (panel b) is the Hoyle-Lyttleton rate (Eq. 1 in paper I), solid are the Bondi-Hoyle approximation formula (Eq. 3 in paper I) and half that value. The upper solid bold curve represents the numerically calculated mass accretion rate. The lower three curves of the left panels trace the x (bold), y (medium) and z (thin) component of the angular momentum accretion rate. The components are multiplied by the factor written down near the curves in order to amplify the deflections of the curves.

| Model | $\mathcal{M}_\infty$ | $L$ ($R_A$) | $R$ ($R_A$) | $g$ | $\delta$ ($R_A$) | $\epsilon$ | $t_f$ | $\overline{\dot M}$ ($\dot M_{BH}$) | $\widehat{\dot M}$ ($\dot M_{BH}$) | $S$ ($\dot M_{BH}$) | $P$ | $t_u$ | $d_{ss}$ ($R_A$) | $s$ ($\mathcal{R}$) | $l_{rms}$ ($R_A c_\infty$) |
|---|---|---|---|---|---|---|---|---|---|---|---|---|---|---|---|
| AS | 0.6 | 32. | 0.02 | 9 | 1/256 | 3 | 8.43 | 1.18 | 1.20 | 0.01 | $>t_f$ | $>t_f$ | - | 0.17 | 0.001 |
| AM | 0.6 | 32. | 0.1 | 7 | 1/64 | 4 | 16.9 | 1.72 | 1.74 | 0.01 | $>t_f$ | $>t_f$ | - | 0.13 | 0.001 |
| AL | 0.6 | 128. | 1. | 6 | 1/8 | 5 | 68.1 | 38.7 | 38.7 | 0.05 | $>t_f$ | $>t_f$ | - | 0.16 | 0.001 |
| BS | 1.4 | 32. | 0.02 | 9 | 1/256 | 3 | 19.5 | 2.10 | 2.12 | 0.01 | $>t_f$ | $>t_f$ | 0.07 | 0.94 | 0.001 |
| BM | 1.4 | 32. | 0.1 | 7 | 1/64 | 4 | 28.2 | 2.13 | 2.14 | 0.01 | $>t_f$ | $>t_f$ | 0 | 0.785 | 0.001 |
| BL | 1.4 | 128. | 1. | 6 | 1/8 | 5 | 113. | 5.68 | 5.71 | 0.01 | $>t_f$ | $>t_f$ | -1.0 | 0.313 | 0.001 |
| CS | 3.0 | 32. | 0.02 | 9 | 1/256 | 3 | 9.32 | 0.98 | 1.29 | 0.14 | vari | 3.5 | 0.1-0.3 | 5.5 | 0.072 |
| CM | 3.0 | 32. | 0.1 | 7 | 1/64 | 4 | 20.8 | 1.04 | 1.32 | 0.11 | vari | 4.0 | 0–0.2 | 4.0 | 0.099 |
| CL | 3.0 | 128. | 1. | 6 | 1/8 | 5 | 100. | 2.74 | 2.74 | 0.01 | 8.3 | $>t_f$ | -1.0 | 0.289 | 0.001 |
| DS | 10. | 32. | 0.02 | 9 | 1/256 | 3 | 2.84 | 0.73 | 1.06 | 0.11 | vari | 0.6 | 0.06-0.2 | 11.2 | 0.249 |
| DM | 10. | 32. | 0.1 | 7 | 1/64 | 4 | 5.26 | 0.76 | 0.86 | 0.05 | vari | 0.5 | 0–0.2 | 10.5 | 0.363 |
| DL | 10. | 128. | 1. | 6 | 1/8 | 5 | 34.2 | 2.26 | 2.28 | 0.01 | $>t_f$ | $>t_f$ | -2.0 | 0.231 | 0.002 |

**Table 1.** Parameters and some computed quantities for all models. $\mathcal{M}_\infty$ is the Mach number of the unperturbed flow, $L$ size of the largest grid, $R$ radius of the accretor, $g$ = number of grid nesting depth levels, $\delta$ = the size of one zone on the finest grid, $\epsilon$ is the softening parameter (zones) for the potential of the accretor (see paper I), $t_f$ is the total time of the run, $\overline{M}$ is the integral average of the mass accretion rate, $\widehat{M}$ is the maximum mass accretion rate, $S$ is one standard deviation around the mean $\overline{M}$ of the mass accretion rate fluctuations, $\dot{M}_{BH}$ is defined in Eq. 3 of paper II, $P$ is the approximate period of the accretion rate fluctuations ("vari": no periodicity is visible), $t_u$ is the approximate time at which the flow becomes unstable (the time at which the ang. mom. accretion rate reaches a value of 0.6), $d_{ss}$ approximate shock standoff distance (measured from the center of the accretor, negative:downstream, zero: shock does not clear the surface), $s$ is the entropy (Eq. 4 in paper II), $l_{rms}$ is the root mean square of the specific angular momentum magnitude; all time units are $R_A/c_\infty$; all models have a specific heat ratio of $\gamma = 4/3$, the initial density distribution is randomly perturbed by 3%, and the number $N$ of zones per grid dimension is 32.

The axisymmetric steady state flow is characterised by very small accretion rates of all angular momentum components: the contributions of antipodal parts of the accretor surface cancel.

### 3.2. Mildly supersonic accretors

As soon as the relative velocity between accretor and medium is supersonic a shock front develops: contour plots of the density of models BL, BM and BS can be seen in Fig. 2. In all these models the accretor moves at a velocity of Mach 1.4 relative to the surrounding medium. The shape of the accretion cone on larger scales is qualitatively very similar to what has already been shown for models KS, KM and KL ($\gamma = 5/3$) in paper III. We do not repeat the plots here. However, the flow in the direct vicinity of the accretor, differs markedly when reducing the size of the accretor. While the models with very large accretors (KL from paper III and BL) do not differ (compare panel 2a shown here with panel 4b from paper III), the models with medium sized accretors (KM and BM) do show a difference: the density in the vicinity of the accretor is higher in models with smaller $\gamma$ and the bow shock stand off distance from the surface is much smaller. Finally, the models BS and KS show the largest difference: in model KS the inflow is approximately spherically symmetric out to distances of about 0.2 accretion radii and the shock front is at 0.4 $R_A$, while in model BS the distance of the bow shock is less than 0.1 $R_A$.

Additionally, the bow shock is not completely convex, but shows a concave form in regions closer to the accretor than about 0.2 $R_A$. A magnification of the central region of panel 2e can be seen in Fig. 3. This figure shows the divergence of the velocity field as shades of gray. The divergence of the velocity is a good indicator for shocks, since in shocks matter is decelerated and thus a high negative value of the divergence appears. These higher negative values are shown by darker shades. Additionally to the concave bow shock one can discern two dark lines connecting the bow shock to the accretor. These lines are a slice of a ring-shaped shock, the axis of the ring being the x-axis.

These features have been described by Hunt (1979), and so we confirm the existence also in three dimensional models. However, we observed them only in model BS. The A models are subsonic so no shocks form in the first place, the models BM and BL as well as CL and DL have too large accretors, and finally the models CM, CS, DM and DS do not reach a steady state. It has also not been seen in the $\gamma = 5/3$ model KS.

The accretion rates of mass and angular momentum can be seen in Fig. 2 panels b, d and f. They all show that the models tend toward a steady state, in which the mass accretion rate monotonously approaches some maximum value (which is listed in Table 1) and the angular momentum approaches zero.

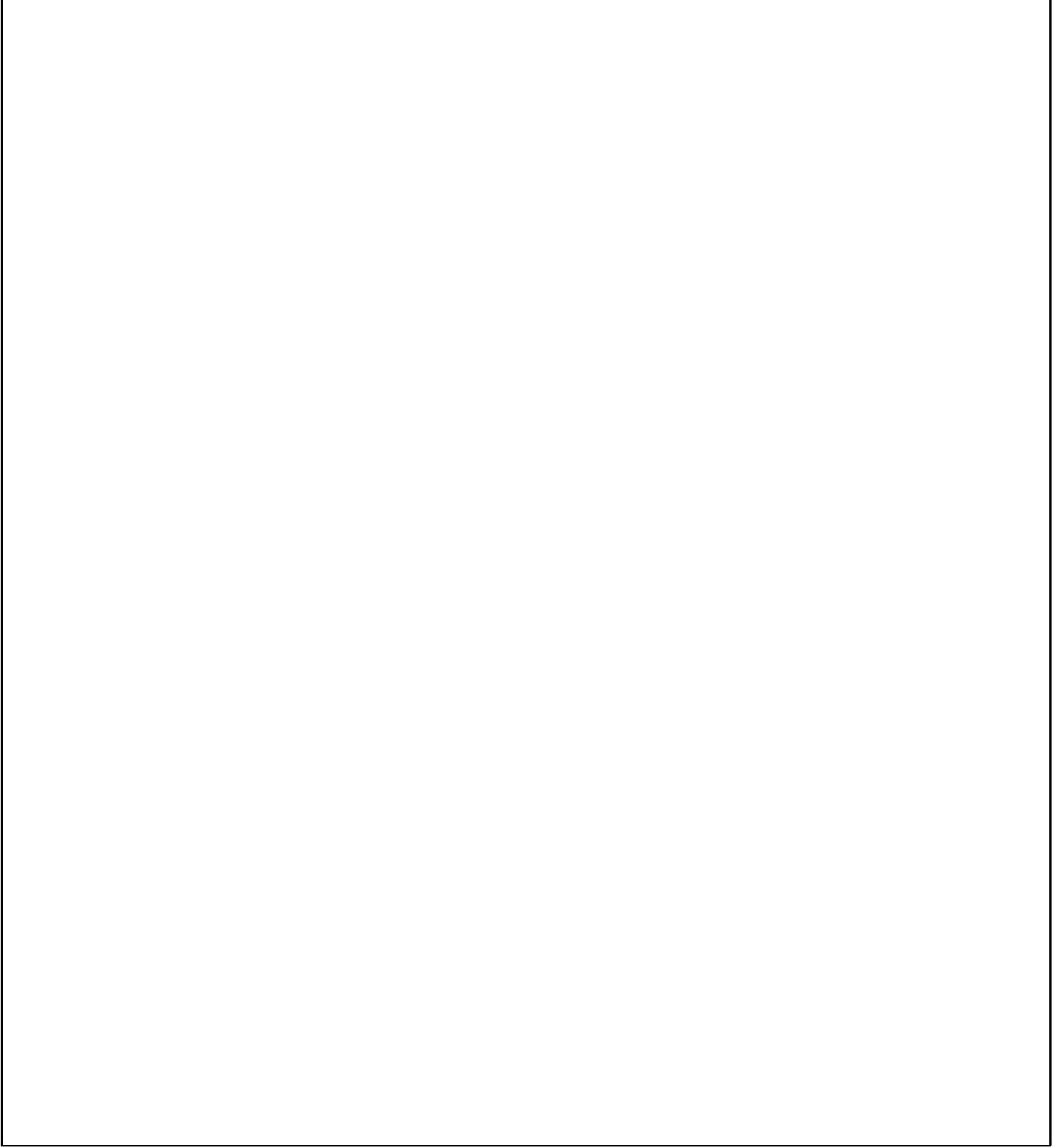

**Fig. 2.** Panels a, c and e are contour plots showing snapshots of the density together with the flow pattern in a plane containing the center of the accretor for all mildly supersonic models BL (panel a), BM (panel c), and BS (panel e). The contour lines are spaced logarithmically in intervals of 0.05 dex for models BL and 0.1 dex for models BM and BS. The bold contour levels correspond to $\log \rho = 0.01$ and $\log \rho = 0.3$ for model BL, $\log \rho = 0.01$ and $\log \rho = 1.0$ for model BM, and $\log \rho = 1.0$ for model BS. The time of the snapshot together with the velocity scale is given in the legend at the upper right hand corner in all three panels. The accretion rates of several quantities are plotted as a function of time for all mildly supersonic models BL, BM and BS in panels b, d and f. The straight horizontal lines show the analytical mass accretion rates: dotted (panel b) is the Hoyle-Lyttleton rate (Eq. 1 in paper I), solid are the Bondi-Hoyle approximation formula (Eq. 3 in paper I) and half that value. The upper solid bold curve represents the numerically calculated mass accretion rate. The lower three curves of the left panels trace the x (bold), y (medium) and z (thin) component of the angular momentum accretion rate. The components are multiplied by the factor written down near the curves in order to amplify the deflections of the curves. See also caption of Fig. 1.

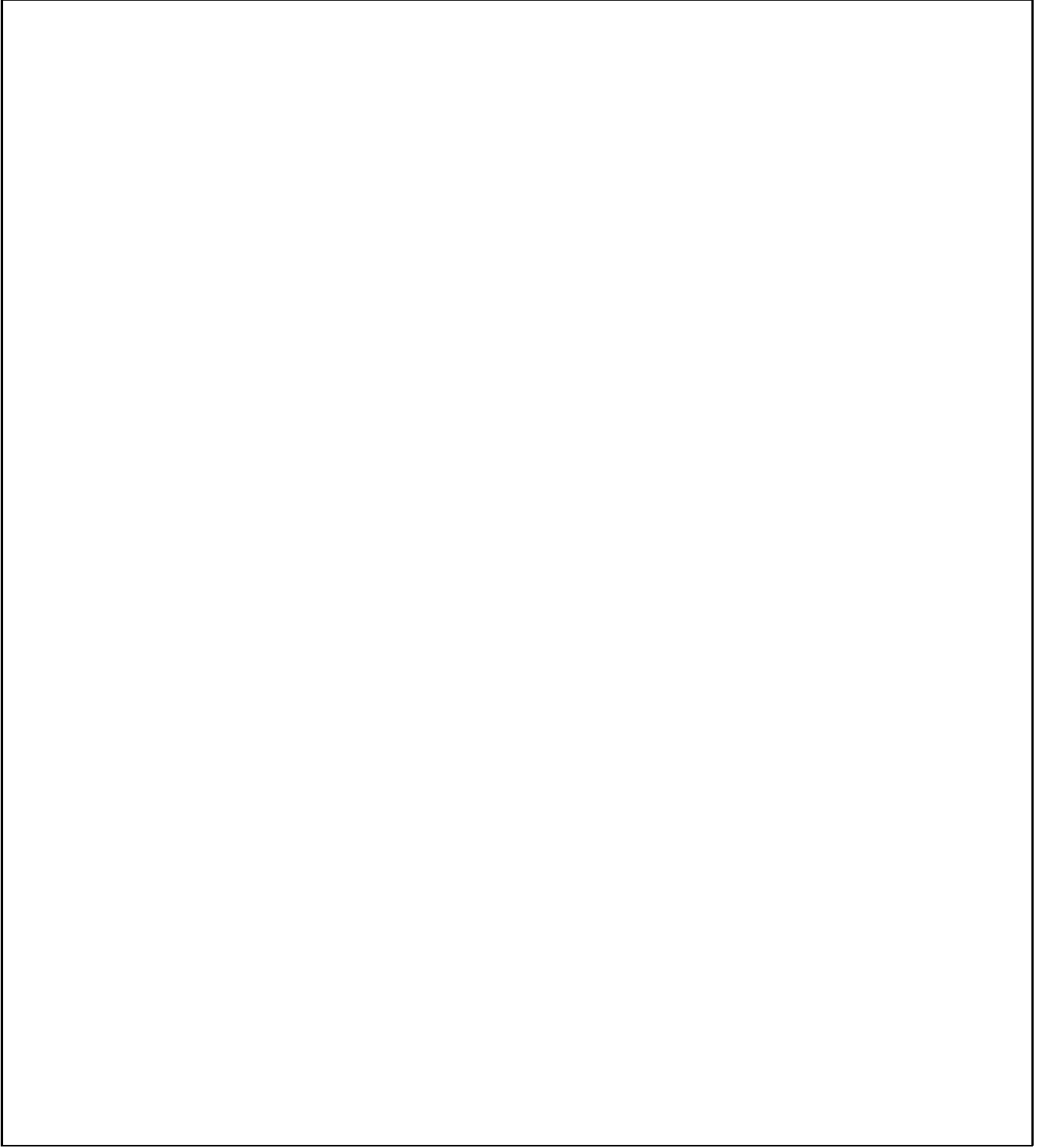

**Fig. 4.** Panels a, c and e are contour plots showing snapshots of the density together with the flow pattern in a plane containing the center of the accretor for all moderately supersonic models CL (panel a), CM (panel c), and CS (panel e). The contour lines are spaced logarithmically in intervals of 0.1 dex. The bold contour levels correspond to $\log \rho = 0.01$ and $\log \rho = 1$. The accretion rates of several quantities are plotted as a function of time for all moderately supersonic models CL, CM and CS in panels b, d and f. The upper solid bold curve represents the numerically calculated mass accretion rate. The lower three curves of the left panels trace the x (bold), y (medium) and z (thin) component of the angular momentum accretion rate. See also caption of Fig. 1.

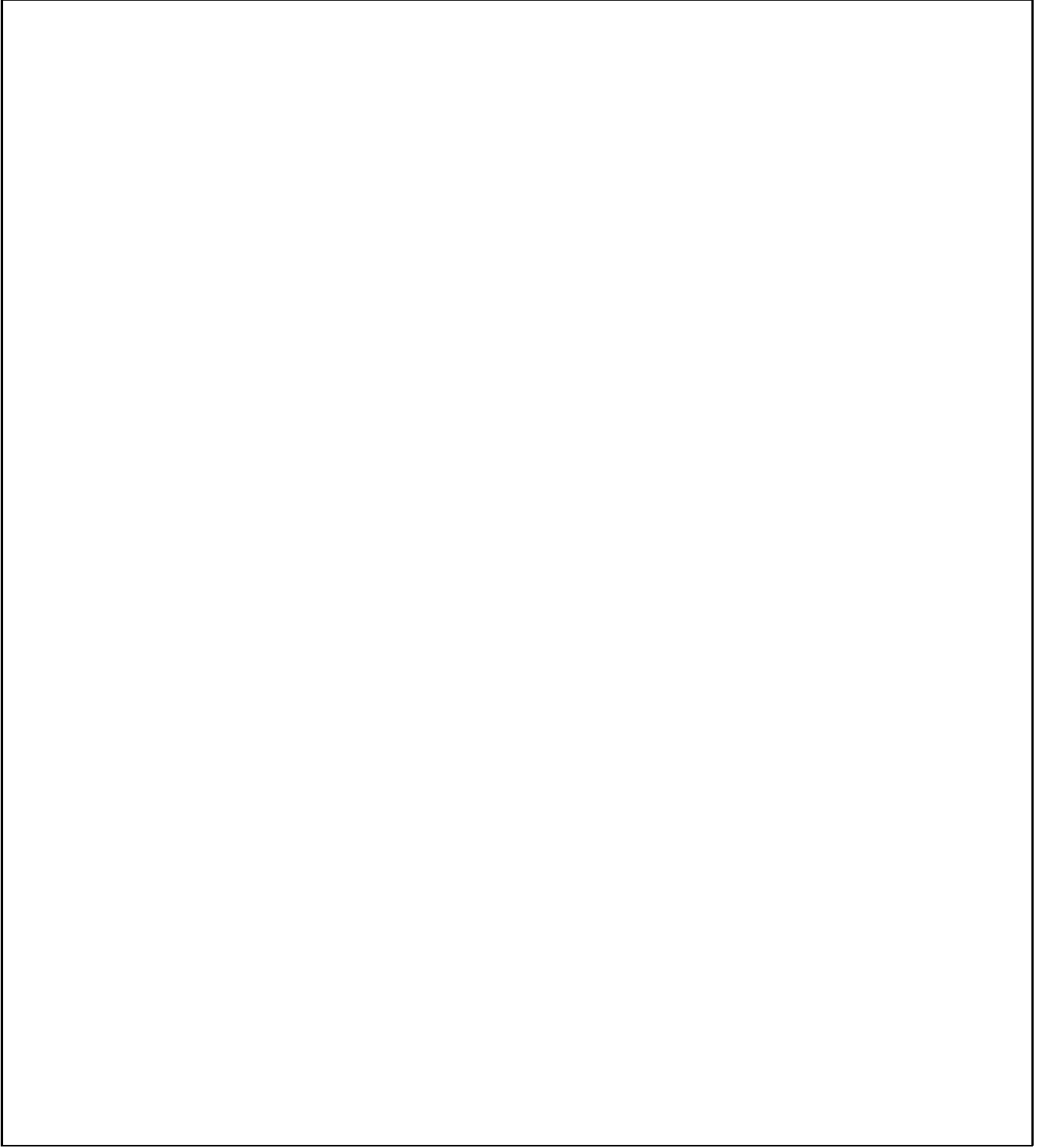

**Fig. 5.** Panels a, c and e are contour plots showing snapshots of the density together with the flow pattern in a plane containing the center of the accretor for all highly supersonic models DL (panel a), DM (panel c), and DS (panel e). The contour lines are spaced logarithmically in intervals of 0.1 dex. The bold contour levels correspond to $\log \rho = 0.01$ and $\log \rho = 1$. The accretion rates of several quantities are plotted as a function of time for all highly supersonic models DL, DM and DS in panels b, d and f. The upper solid bold curve represents the numerically calculated mass accretion rate. The lower three curves of the left panels trace the x (bold), y (medium) and z (thin) component of the angular momentum accretion rate. See also caption of Fig. 1.

**Fig. 3.** A magnification of the central region of model BS (cf. Fig. 2e) showing the divergence of the velocity in the xy-plane. Darker tones represent higher values of the negative divergence and thus stronger compression of matter. These are the places of shocks. The circular black shape around (x,y)=(0,0) is the accretor, the dark band running roughly vertically at x=0.08 is the bow shock. The black line at x=0.06 is a numerical artefact due to the interface between a coarse and a fine grid. Note an additional ring-shaped shock connecting the accretor and the bow shock, of which only a slice is visible in this plot, which reduces the ring to two lines.

## 3.3. Moderately and highly supersonic accretors

We describe the 'C'-models ($\mathcal{M}_\infty = 3$) and the 'D'-models ($\mathcal{M}_\infty = 10$) together, since they differ only quantitatively, not qualitatively.

### 3.3.1. Large accretors, $R = 1 R_A$

The contour plot of the density distribution of models CL and DL displayed in Figs. 4a and 5a show typical features that have have already been seen for models H6 and FL in papers II and III, respectively. A tail shock develops and directly above the downstream surface of the accretor, a dip in the density distribution is present. The gravitational force at distances larger than the accretion radius is so weak that the surrounding medium is hardly deflected. However, the large geometrical cross section does absorb material and the medium has to fill the "hole" in the wake of the moving accretor. Most of the shock cone is of course filled with matter of higher density ($\rho > 1$).

In Figs. 4b and 5b we show how the accretion rates of several quantities evolve with time. Model DL tends toward a steady state, while model CL shows a very small periodic fluctuation of the angular momentum components. This is traceable to two zones directly at the surface of the accretor, which cycle through exchanging their values of density and temperature. 

the cycle leads us to believe that it might be a numerical artefact. It does not seem to have any other detrimental repercussions.

### 3.3.2. Small accretors

The four models CM, CS, DM and DS are the only models listed in Table 1 that did not converge to a quiescent steady state but showed an unstable, irregularly fluctuating, flow pattern. This type of flow has already been described in papers II and III. In Figs. 4c and 4e as well as in Figs. 5c and 5e snapshots of the typical density distribution have been plotted for these four models. They show the unstable flow patterns resulting in non-spherical accretion flow. Lower density pockets, vortices etc. can be discerned.

In models CS and DS, which contain a very small accretor, (with radius $R = 0.02 R_A$) the shock front clears the surface of the accretor and stays well away from it (approximately to $10R = 0.2 R_A$), However, in the models CM and DM (with $R = 0.1 R_A$) the shock front ends at the surface of the accretor most of the time. Only sporadically does a dome shaped bubble appear (cf. Matsuda et al, 1992; Koide et al, 1991) surrounding the accretor. This difference is not so pronounced in the $\gamma = 5/3$-models (compare e.g. models FM and FS in paper III, Fig. 8c and 8d). In the $\gamma = 4/3$-models we present here, the equation of state is softer and thus the pressure in the immediate vicinity lower than for the $\gamma = 5/3$-models. So the shock front is not supported as well and resides nearer to the surface of the accretor.

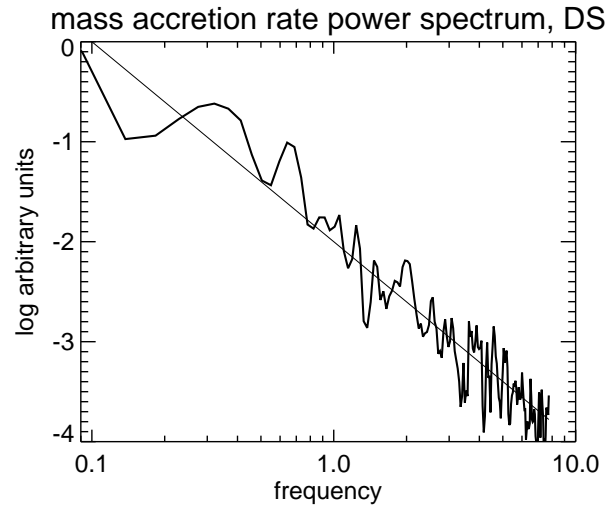

**Fig. 6.** The power spectrum of the mass accretion rate of model DS is shown. Frequency has the reciprocal units of time, i.e. $c_\infty / R_A$. The thin straight line shows the relation $P \propto f^{-2}$.

The unsteady flow reflects itself in the fluctuating accretion rates also shown in Figs. 4 and 5 panels d and f. The temporal evolution of all accretion rates plotted (mass, angular momen-apparent after an initial phase in which the flow builds up. Only model DS shows a slight periodicity which can better be seen when a Fourier transform of the mass accretion rate is done (Fig. 6). A local maximum is visible around a frequency of 3.2, with its harmonic at around 6.5. Thus the period is about 0.3 time units. This as well as the other models show a general run of the spectrum proportional to $f^{-2}$, characteristic of a random-walk noise.

Because of all the unstable flow we do not see the features reported by Hunt (1979), who mentions an indented bow shock and the presence of a secondary shock. These have sizes of about $0.03R_A$, so even if these features did appear in our simulations they would be dissolved by the turbulent flow patterns.

## 4. Analysis of Results

### 4.1. Mass accretion rate

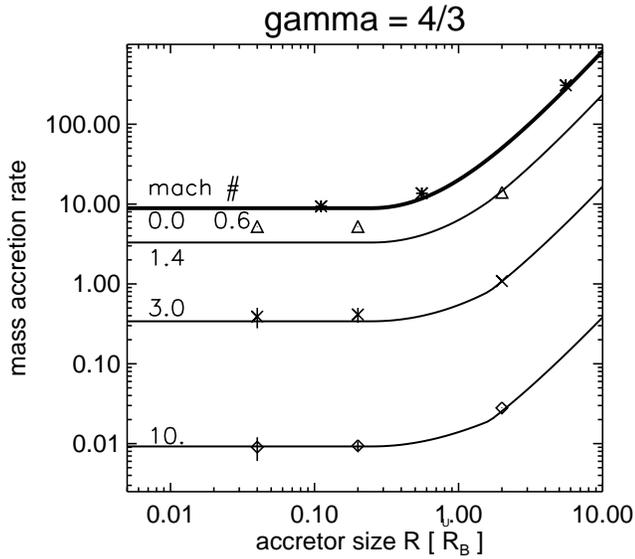

**Fig. 7.** Mass accretion rates (units: $4\pi R_B^2 c_\infty \rho_\infty$) as a function of accretor size (units: $\check{R}_B$ (Eq. 7 in Paper III), *not* accretion radii). The Mach number is an additional parameter and is specified by the numbers to the right of the y-axis. The topmost bold curve corresponds to a Mach number of zero, i.e. a stationary accretor. The other lines belong to Mach numbers of 0.6, 1.4, 3.0 and 10., whereby the curve of Mach 0.6 completely overlaps with the curve belonging Mach 0. The numerical results to the corresponding Mach numbers are plotted with different symbols: the stars (∗), triangles (△) crosses (×) and diamonds (◇) are the results of models with $\mathcal{M}_\infty$=0.6, 1.4, 3.0 and 10, respectively. The "error bars" extending from the symbols indicate two standard deviations from the mean ($S$ in Table 1), for models in which the mass accretion rate fluctuated.

All mass accretion rates obtained in this work are collected in Fig. 7. Note that in this figure the unit of length (for the accretor size) is $\check{R}_B$ (Eq. 7 in Paper III), *not* $R_A$. In Paper III we had of Bondi, 1952) the mass accretion rates found in the previous papers I, II and III. Since all those previous models were done using a specific heat ratio of 5/3, the formula is adapted only to that one $\gamma$. However, it does include correctly the limit case of spherical Bondi accretion, i.e. no relative bulk motion between medium and accretor. In Fig. 7 we included the mass accretion rate that follows from the interpolation formula for $\gamma = 4/3$ as solid curves.

Two points seem noteworthy. (a) Since the sonic point is at a distance of 1/4 $R_a$ the curves run horizontally for all accretors that have a radius smaller than this value of 1/4. This is due to the fact that within the sonic radius the flow is supersonic and thus a change in size of the accretor (as long as this radius stays smaller than the sonic distance) should not influence the accretion flow. In the theory derived by Bondi (1952) for spherically symmetric accretion the maximum accretion rate is fixed by the state at the sonic point. Within numerical accuracy, also the maximum accretion rates derived numerically are equal for those models in which the accretor size is smaller than the sonic distance, e.g. the mass accretion rate for models DM and DS are 0.73 and 0.76, respectively (cf. Table 1). (b) The interpolation formula does not fit the mass accretion rates for the models BM and BS (i.e. $\mathcal{M}_\infty = 1.4$) too well: the numerical values lie above the interpolation curves. Before we alter the formula in any way to accomodate this fact, we will wait for the results of the simulations in which we set $\gamma = 1.01$ (which we have already begun) in order to get a better overview on the influence that $\gamma$ has.

On the whole, the flow patterns are less violent in the $\gamma = 4/3$ models compared to the $\gamma = 5/3$ models. Fig. 8 shows the relative mass fluctuations, i.e. the standard deviation devided by the average mass accretion rate $\overline{\dot{M}}/S$ (cf. Table 1), for all pertinent models. One notices, that in nearly all cases the fluctuations for the $\gamma = 5/3$ models are larger than for those models with $\gamma = 4/3$, when other parameters are kept constant. That the fluctuations of model DM are so much smaller than those of model CM is due to the fact that in model DM the shock front hardly ever clears the surface of the accretor and when it does, it does not venture very far.

Fig. 9 shows three effects that increase the maximum density that is reached on the numerical grid: (1) models with higher Mach numbers, (2) models with smaller accretors and (3) models with lower $\gamma$, all have higher densities. All three points are easy to understand: a smaller accretor allows matter to fall into a deeper potential well, thus compressing it more. Higher Mach numbers imply stronger shocks, across which the density jump is then larger. And finally, with smaller $\gamma$ the EOS is softer, i.e. a higher density is needed to obtain some fixed pressure value, and the density jump across a shock is again larger. Some of these features have already been reported by Shima et al (1985).

### 4.2. Angular momentum accretion rate

For many applications the magnitude of the accreted specific angular momentum is of interest. In Fig. 10 we plot for several models the ratio of temporal rms average of the specific angular momentum $l_{rms}$ over Mach number $\mathcal{M}_\infty$ as a function of the accretor size. The values of $l_{rms}$ and $\mathcal{M}_\infty$ can be found in Table 1. Additionally, at the top left of the plot, the diagonal line shows what values are expected analytically if one assumes that a vortex flows with Kepler velocity $V$ around the accretor, just above the surface, i.e. $l_{rms}$ is given by $l_s = RV = \sqrt{R}\mathcal{M}_\infty$. This plot (Fig. 10) enlarges the upper left portion of Fig. 13 in Paper III. So only those models that exhibit unstable, active flow patterns are shown, while all models that converge toward a steady state have values of $l_{rms}/\mathcal{M}_\infty$ around or below $10^{-3}$.

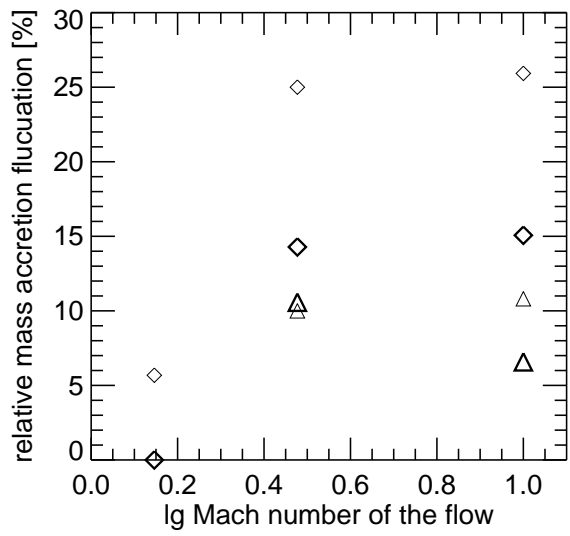

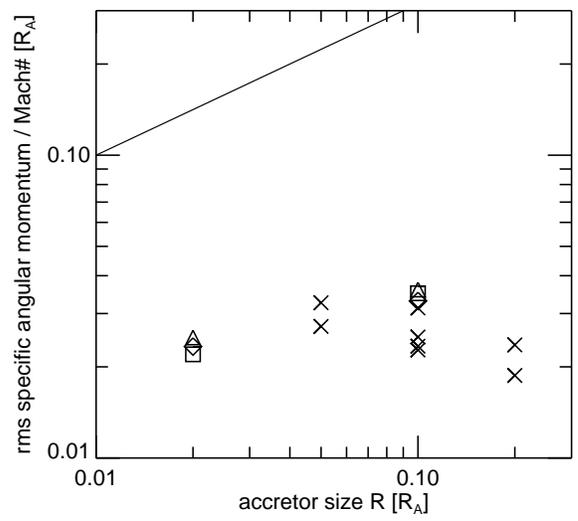

**Fig. 8.** The relative mass fluctuations, i.e. the standard deviation $S$ divided by the average mass accretion rate $\overline{\dot{M}}$ (cf. Table 1), is shown as a function of the Mach number of the flow. Diamonds ($\diamond$) denote models in which the accretor has a radius of 0.02 $R_A$, triangles ($\triangle$) models with 0.1 $R_A$. The large bold symbols belong to models with $\gamma = 4/3$, while smaller thin symbols belong to models with $\gamma = 5/3$.

**Fig. 10.** As a function of accretor size the ratio of temporal rms average of the specific angular momentum $l_{rms}$ over Mach number $\mathcal{M}_\infty$ is plotted for different models: squares ($\square$) for models FS and FM from paper III, crosses ($\times$) for the models from paper II, diamonds ($\diamond$) for models CM and CS, and finally triangles ($\triangle$) for models DM and DS. The diagonal straight line at the top left shows what values $l_s$ (see text) are obtained if the flow would rotate with Kepler velocity along the accretor surface.

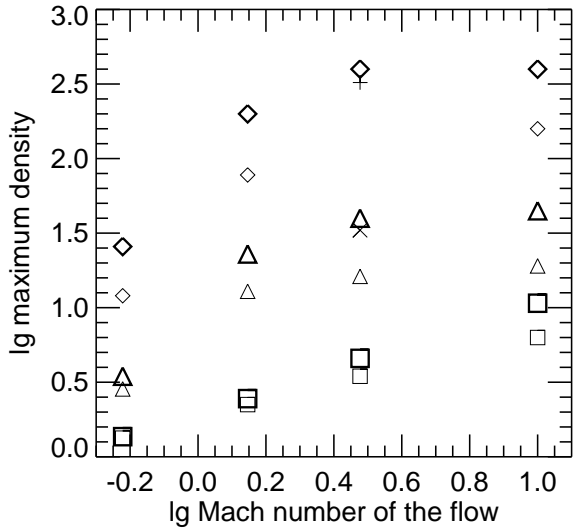

**Fig. 9.** The maximum density that is reached on the grid is shown as a function of the Mach number of the flow. Diamonds ($\diamond$) denote models in which the accretor has a radius of 0.02 $R_A$, triangles ($\triangle$) models with 0.1 $R_A$, squares ($\square$) models with $1.0 R_A$, while the cross ($\times$) is model S6 with $0.2 R_A$ and the plus (+) is model T10 with $0.01 R_A$. The large bold symbols belong to models with $\gamma = 4/3$, while smaller thin symbols belong to models with $\gamma = 5/3$.

Note that the angular momentum accreted in the models presented in this paper ($\gamma = 4/3$) is equal (to within numerical accuracy) to the values produced by the $\gamma = 5/3$ models. So although the mass accretion rates differ in models with differing $\gamma$ this is not the case for the specific angular momentum. It is also interesting to note the following. In a preliminary simulation (Ruffert & Anzer, 1995) we modeled the accretion of a medium with a velocity gradient of 3% per accretion radius distance. In this simulation the average specific angular momentum accreted is 0.097 at Mach 3 for an accretor of size $0.1 R_A$. So it would be placed at y=0.032 in Fig. 10, right on top of the points that have been found for homogeneous media. This is a coincidence, since the accretion rate is surely dependent on the choice of the magnitude of the gradient. However, the form of the dependence is as yet unknown.

**Fig. 11.** The total angular momentum (magnitude of the vector) accretion rate is plotted as a function the mass accretion rate for models CM, CS, DM and DS. Each dot displays the two rates at one moment in time.

In Fig. 11 we plot the angular momentum accretion rate as a function of the mass accretion rate at every second time step. The correlation visible in $\gamma = 5/3$ models (papers II and III), i.e. the accretion of large angular momentum values tends to happen preferentially when the mass accretion rates are low, is not so obvious any longer (in $\gamma = 4/3$ models). Independently of how much angular momentum is accreted, the mass accretion rate remains in a fixed range.

### 4.3. Comparison of shock opening angles

As has already been pointed out (e.g. Petrich et al, 1989), the shock opening angle ($\theta$, angle of the shock front to the direction of the unperturbed flow) seen in the simulations tends to be larger than the values predicted analytically, $\Theta = \arcsin(1/\mathcal{M})$, for large distances from the accretor. Since the shock strength decreases with increasing distance from the accretor it is difficult to accurately quantify an angle from the contour plots shown further above. The angles are plotted in Fig. 12 by vertical bars, the length of which indicate the uncertainty of measurement and the variation of models differing in accretor size. There is a clear trend toward numerical angles being larger than the analytical ones at large Mach numbers. The opening angles for $\gamma = 4/3$ models tend to be slightly smaller than those of the $\gamma = 5/3$ models. Petrich et al (1989, Fig. 12) report a factor of 2 larger numerical angles compared to the analytic ones for $\gamma = 4/3$. We reproduce this factor, however only for the models involving large Mach numbers ($\mathcal{M}_\infty = 10$). Until this question is resolved, one should bare in mind this difference between analytic and numeric angles when interpreting observations of shock cones (e.g. neutron stars in the ISM, etc.).

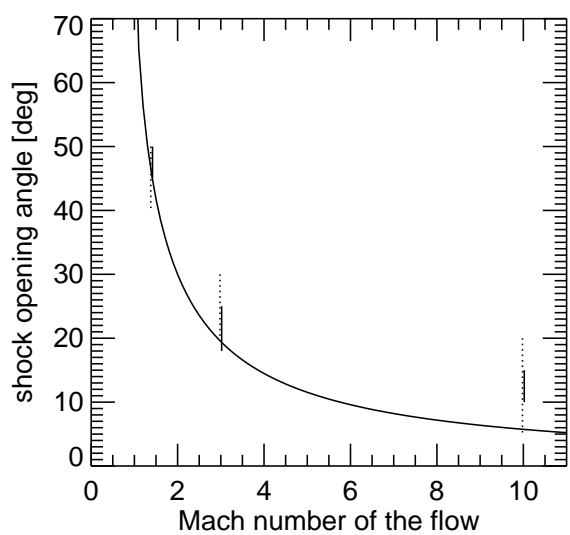

**Fig. 12.** The shock opening angles for various models are plotted versus speed of the flow. The vertical lines show the numerical models, solid represents the $\gamma = 4/3$ models, dashed the $\gamma = 5/3$ models. The curve shows the relation $\Theta = \arcsin(1/\mathcal{M})$.

## 5. Conclusions

We draw the following conclusions from the results presented here:

1. For $\gamma = 4/3$ the flow (and thus the accretion rate of all variables) exhibits active unstable phases only for models in which the flow speed is large enough and the accretor size small enough: all models with Mach numbers of 0.6 or 1.4 tend toward a steady state with quiescent accretion flow, as well as the Mach 3 and 10 models with an accretor size of $1 R_A$. Only the models at Mach 3 and 10 and with accretor sizes 0.1 and 0.02 show unstable flow patterns.
2. The shock front stays closer to the accretor in the $\gamma = 4/3$ models compared to the $\gamma = 5/3$ models. So much so, that in models with an accretor size of $0.1 R_A$ (and a flow velocity of Mach 3 or 10) the shock front hardly ever clears the accretor surface.
3. The model with Mach 0.6 yields accretion rates identical (to numerical accuracy) to those in stationary accretor models.
4. In one mildly supersonic and stable model with very small accretor (0.02 accretion radii) the bow shock is concave at distances up to 0.2 accretion radii from the accretor, and an additional ring shaped shock connects the bow shock to the surface of the accretor.
5. The mass accretion rates are slightly larger for the $\gamma = 4/3$ models compared to $\gamma = 5/3$ and the phenomenological fit improved.
6. The correlation between accreted mass and angular momentum observed in the $\gamma = 5/3$ models (where larger angular momenta are accreted when the mass accretion rate is low) is not visible in the $\gamma = 4/3$ models presented here.
7. The rms of the specific angular momentum accreted does not change when $\gamma$ is varied.

Movies of the dynamical evolution are available in mpeg format
in the WWW at `http://www.mpa-garching.mpg.de/~mor/bhla.html`

*Acknowledgements.* I would like to thank Dr. F. Meyer and Dr. U. Anzer for carefully reading and correcting the manuscript. The calculations were done at the Rechenzentrum Garching.